\documentclass[hidelinks,conference]{IEEEtran}
\IEEEoverridecommandlockouts

\usepackage{cite}
\usepackage{amsmath,amssymb,amsfonts}
\usepackage{graphicx}
\usepackage{textcomp}
\usepackage{dsfont}
\usepackage{xcolor}
\usepackage{orcidlink}
\usepackage{bm}
\usepackage{caption}
\usepackage{subcaption}
\usepackage{relsize} 
\usepackage{outlines} 
\usepackage{soul}
\usepackage{algorithm, algpseudocode}
\usepackage{lipsum}  
\usepackage[acronym,shortcuts]{glossaries}
\usepackage[bottom]{footmisc}

\def\BibTeX{{\rm B\kern-.05em{\sc i\kern-.025em b}\kern-.08em
T\kern-.1667em\lower.7ex\hbox{E}\kern-.125emX}}

\newcommand{\trans}[0]{^{\mathsf{T}}}
\newcommand{\transs}[0]{^{\!\mathsf{T}}}
\newcommand{\transss}[0]{^{\!\!\mathsf{T}}}
\newcommand{\herm}[0]{^{\mathsf{H}}}
\newcommand{\R}[0]{^{\mathrm{R}}}
\newcommand{\I}[0]{^{\mathrm{I}}}

\newcommand{\Exp}[0]{\mathbb{E}}

\newacronym{SM}{SM}{spatial modulation}
\newacronym{GSM}{GSM}{generalized spatial modulation}
\newacronym{QSM}{QSM}{quadrature spatial modulation}
\newacronym{GQSM}{GQSM}{generalized quadrature spatial modulation}
\newacronym{OS-QSM}{OS-QSM}{optimized scalable quadrature spatial modulation}
\newacronym{IQ}{IQ}{in-phase and quadrature}

\newacronym{5G}{5G}{fifth generation}
\newacronym{B5G}{B5G}{beyond fifth generation}
\newacronym{6G}{6G}{sixth generation}
\newacronym{mmWave}{mmWave}{millimeter-wave}
\newacronym{THz}{THz}{Terahertz}
\newacronym{MIMO}{MIMO}{multiple-input multiple-output}
\newacronym{mMIMO}{mMIMO}{massive multiple-input multiple-output}
\newacronym{XL-MIMO}{XL-MIMO}{extra-large MIMO}
\newacronym{MU}{MU}{multi-user}
\newacronym{P2P}{P2P}{point-to-point}
\newacronym{AWGN}{AWGN}{additive white Gaussian noise}
\newacronym{CSI}{CSI}{channel state information}
\newacronym{SotA}{SotA}{state-of-the-art}
\newacronym{RF}{RF}{radio frequency}
\newacronym{RIS}{RIS}{reconfigurable intelligent surface}
\newacronym[\glslongpluralkey={spectral efficiencies}]{SE}{SE}{spectral efficiency}
\newacronym[\glslongpluralkey={energy efficiencies}]{EE}{EE}{energy efficiency}

\newacronym{STBC}{STBC}{space-time block code}
\newacronym{MQAM}{$M$-QAM}{$M$-ary quadrature amplitude modulation}
\newacronym{STC}{STC}{space-time coding}
\newacronym{IM}{IM}{index modulation}

\newacronym{ML}{ML}{maximum likelihood}
\newacronym{SD}{SD}{sphere decoding}
\newacronym{MSE}{MSE}{mean squared error}
\newacronym{BER}{BER}{bit error rate}
\newacronym{SNR}{SNR}{signal-to-noise ratio}
\newacronym{IC}{IC}{interference cancellation}
\newacronym{MMSE}{MMSE}{minimum mean-squared-error}
\newacronym{CS}{CS}{compressive sensing}
\newacronym{CPU}{CPU}{central processing unit}
\newacronym{UVD}{UVD}{unit vector decomposition}
\newacronym{ROMP-OSIC}{ROMP-OSIC}{relaxed orthogonal matching pursuit-based ordered successive IC}

\newacronym{MP}{MP}{message passing}
\newacronym{FG}{FG}{factor graph}
\newacronym{BP}{BP}{belief propagation}
\newacronym{GaBP}{GaBP}{Gaussian belief propagation}
\newacronym{SGA}{SGA}{scalar Gaussian approximation}
\newacronym{CLT}{CLT}{central limit theorem}
\newacronym{PDF}{PDF}{probability density function}
\newacronym{GB-ISTA}{GB-ISTA}{greedy boxed iterative soft-thresholding algorithm}
\newacronym{PMF}{PMF}{probability mass function}
\newacronym{i.i.d.}{i.i.d.}{independent and identically distributed}

\newacronym{IER}{IER}{index vector error rate}
\newacronym{MFB}{MFB}{matched filter bound}

\title{Enabling Energy-Efficiency in Massive-MIMO: \\
A Scalable Low-Complexity Decoder for Generalized Quadrature Spatial Modulation \vspace{-0.25ex}}

\author{
Hyeon Seok~Rou, \IEEEmembership{Graduate Student Member,~IEEE},
Giuseppe~Thadeu~Freitas~de~Abreu, \IEEEmembership{Senior Member, IEEE}, \\
David~Gonz{\'a}lez~G.,~\IEEEmembership{Senior Member,~IEEE,} and Osvaldo~Gonsa \vspace{-2ex}
\thanks{\noindent H.~S.~Rou and G.~T.~F.~Abreu are with the School of Computer Science and Engineering, Constructor University, Campus Ring 1, 28759, Bremen, Germany (e-mails: [h.rou, g.abreu]@constructor.university).}
\thanks{D.~Gonz{\'a}lez~G. and O.~Gonsa are with Wireless Communications Technologies, Continental AG, 
65936 Frankfurt/Main, Germany (e-mails: david.gonzalez.g@ieee.org, osvaldo.gonsa@continental-corporation.com).}}

\begin{document}
\maketitle

\begin{abstract}
\Acf{GQSM} schemes are known to achieve high energy- and spectral-~efficiencies by modulating information both in transmitted symbols and in coded combinatorial activations of subsets of multiple transmit antennas.
A challenge of the approach is, however, the decoding complexity which scales with the efficiency of the scheme.
In order to circumvent this bottleneck and enable high-performance and feasible \acs{GQSM} in \ac{mMIMO} scenarios, we propose a novel decoding algorithm which enjoys a complexity order that is independent of the combinatorial factor.
This remarkable feature of the proposed decoder is a consequence of a novel vectorized \ac{GaBP} algorithm, here contributed, whose \ac{MP} rules leverage both pilot symbols and the \ac{UVD} of the \acs{GQSM} signal structure.
The effectiveness of the proposed \ac{UVD}-\ac{GaBP} method is illustrated via computer simulations including numerical results for systems of a size never before reported in related literature (up to 32 transmit antennas), which demonstrates the potential of the approach in paving the way towards high energy and spectral efficiency for wireless systems in a truly \ac{mMIMO} setting.
\end{abstract}
\glsresetall

\glsresetall
\glsunset{mMIMO}
\section{Introduction}
\label{sec:introduction}
\vspace{-0.1ex}

\Ac{SM} techniques \cite{Mesleh_TVT08} have been widely studied as a promising \ac{MIMO} transmission scheme, which exploits multiple-antenna resources not only to gain spatial diversity, but also to encode ``\textit{energy-free}'' (spatially-modulated) information in the form of a sparse activation of small subsets of the total available transmit antennas.
{\color{black} This unique feature of \ac{SM} schemes allow large data to be encoded with lower effective transmit energy.

Therefore, \ac{SM}} is an excellent candidate enabling technology for \ac{B5G} and \ac{6G} wireless communication systems \cite{Zhang_VTM19} because it offers simultaneously: high gains in \ac{EE} and \ac{SE}; a sparse utilization of \ac{RF} chains, which are highly appealing to \ac{mmWave} and \ac{THz} systems \cite{Rappaport_Access19}; and great flexibility, since the symbols actually transmitted are not directly relevant to the antenna subset activation scheme itself.

Motivated by the aforementioned advantages, a rapid development of enhanced \ac{SM} designs aiming to maximize energy- and spectral-efficiencies, and minimize error rates and decoding complexities can be found in recent literature \cite{Mesleh_TVT14, Basar_TC11, Rou_TWC22, Wang_TVT20, Li_WPC17, An_TVT22, Rou_Asilomar22_QSM}.
An excellent example is the \ac{QSM} design \cite{Mesleh_TVT14}, in which the \ac{SM} technique is applied independently to each \ac{IQ} component of the transmit symbols, resulting not only a two-fold increase in the rate of the spatially-modulated information, but also in improved \ac{BER} performance compared to the original \ac{SM} scheme of \cite{Mesleh_TVT08}.
Further examples in this line of work are the improved \ac{GQSM} designs incorporating \acp{STC} and other techniques to optimize the antenna activation patterns, as proposed in \cite{Basar_TC11, Rou_TWC22}.

Since naive decoding of \ac{GQSM} requires searching over a combinatorial space of size $\binom{\alpha N_T}{P}{~\!\;\!\!}^{\!2} \!\cdot\! M^P$ -- where $N_T$ is the number of transmit antennas, $\alpha$ is the codebook amplification factor ($\alpha \!=\! 1$ for the classic \ac{GQSM}), $P$ is the number of symbols transmitted, and $M$ is the size of the complex symbol constellation -- another line of work towards improving \ac{GQSM} is to lower decoding complexity.
One example is the \ac{CS}-based method of \cite{Rou_TWC22}, which can achieve a significantly low complexity, but may suffer from high spatial domain error since the elaborate \ac{GQSM} codebook structure cannot be incorporated in the estimation rules.
On the other hand, the \ac{SD} approach of \cite{Wang_TVT20} and the reduced-search methods of \cite{Li_WPC17,An_TVT22} were shown to approach \ac{ML} performance, but can only reduce the squared-combinatorial complexity order by a linear factor.

Recently, we proposed a near-\ac{ML} \ac{MP} method for \ac{GQSM} decoding \cite{Rou_Asilomar22_QSM}, which was shown to eliminate the quadratic exponent in the combinatorial factor by employing a novel decomposition of the \ac{GQSM} signal into two independent vectors.
Following the latter work, we propose in this article a novel \ac{MP}-based \ac{GQSM} decoder based on a further decomposition of the \ac{GQSM} signal model into independent unit vectors, which results in a complexity that is completely independent of the combinatorial space, while still incorporating the information of the structured \ac{GQSM} codebook patterns.

The contributions of this article can be summarized as: \vspace{-0.14ex}
\begin{itemize}
\item A novel pilotted \ac{GQSM} decoding algorithm is proposed, which achieves a combinatorial-free complexity order,
\item Tailored \ac{MP} rules are derived based on a \ac{UVD} of the \ac{GQSM} signal model,
\item Simulation results are provided at unprecedented \ac{GQSM} \ac{mMIMO} scales, including system sizes up to $N_T \!=\! 32$.
\end{itemize}



\section{\ac{GQSM} System Model}
\label{sec:system_model}
\vspace{1ex}

\subsection{Transmit Signal Model}

Given a \ac{MIMO} transmitter equipped with $N_T$ antenna elements, the \ac{GQSM} transmit signal is described by \vspace{-0.5ex}
\begin{equation}
\mathbf{x} = \mathbf{x}\R + j\mathbf{x}\I \in \mathbb{C}^{N_T \times 1},
\label{eq:GQSM_transmit_signal}
\vspace{-0.5ex}
\end{equation}
where $\mathbf{x}\R \in \mathbb{C}^{N_T \times 1}$ and $\mathbf{x}\I \in \mathbb{C}^{N_T \times 1}$ are respectively the real and imaginary parts of the \ac{GQSM} transmit signal vector, respectively defined as 
\vspace{-1ex}
\begin{subequations}
{
\color{black}
\label{eq:IQ_GQSM_parts}
\begin{eqnarray}
&\mathbf{x}\R \!\!\!\!&= [0, \hspace{-3ex}\overbrace{s\R_1}^{k\R_1\text{-th position}}\hspace{-3.5ex}, 0, ~\;\!\cdots\!\;~, 0, \hspace{-3ex}\overbrace{s\R_p}^{k\R_p\text{-th position}}\hspace{-3ex}, 0\;, ~\;\!\cdots\!\;~, 0, \hspace{-3ex}\overbrace{s\R_P}^{k\R_P\text{-th position}}\hspace{-3ex}, 0]\trans,~~ \\
&\mathbf{x}\I \!\!\!\!&= [0,\cdots\hspace{-0.15ex}, 0, \hspace{-3ex}\underbrace{s\I_1}_{k\I_1\text{-th position}}\hspace{-2.7ex}, 0, \cdots\!, 0, \hspace{-3ex}\underbrace{s\I_p}_{k\I_p\text{-th position}}\hspace{-2.7ex}, 0,\cdots\!,0, \hspace{-3ex}\underbrace{s\I_P}_{k\I_P\text{-th position}}\hspace{-3ex}, 0]\trans,~~
\end{eqnarray}
}
which contain the real and imaginary parts of the $P$ transmit symbols\footnote{Note that the classic \ac{QSM} \cite{Mesleh_TVT14} is the basic case of the \ac{GQSM} with $P\!=\!1$.} $s_p \!\triangleq\! s\R_p + js\I_p \!\in\! \mathbb{C}$, with $p \in \{1, \cdots\!, P\}$, selected from a discrete constellation $\mathcal{S}$ of cardinality $M$, where the positions of the symbol components are respectively described by the index vectors $\mathbf{k}\R \triangleq [k\R_1, \cdots, k\R_p, \cdots, k\R_P]$ and $\mathbf{k}\I \triangleq [k\I_1, \cdots, k\I_p, \cdots, k\I_P]$.
\end{subequations}

It is important to note that the positions of the symbols between $\mathbf{x}\R$ and $\mathbf{x}\I$ are independent, and that the position of a symbol component in the vector corresponds directly to the activated antenna element at the transmitter.
By exploiting the possible combinatorial patterns of the symbol component positions, the \ac{GQSM} scheme conveys information not only from the encoding of $P$ symbols from $\mathcal{S}$, but also from the selection of $P$ positions out of $N_T$ possible positions, respectively for both $\mathbf{x}\R$ and $\mathbf{x}\I$.

In light of the above, the total information conveyed by the \ac{GQSM} signal $\mathbf{x}$ is given by
\begin{equation}
B_\mathrm{GQSM} = 2B_\mathrm{Sp} + B_\mathrm{Dg} = 2\big\lfloor\!\log_2\!\mathsmaller{\binom{N_T}{P}}\!\big\rfloor + P\log_2(M),
\label{eq:GQSM_bits}
\end{equation}
where $B_\mathrm{Sp}$ is the number of bits spatially encoded by the antenna position selection of $P$ symbol parts, and $B_\mathrm{Dg}$ is the number of bits \textit{digitally} encoded by the transmit symbols.

\subsection{Received Signal Model}

The received signal vector $\mathbf{y} \in \mathbb{C}^{N_R \times 1}$ at the \ac{MIMO} receiver equipped with $N_R$ antennas, is described by \vspace{-0.5ex}
\begin{equation}
\mathbf{y} = \mathbf{H}\mathbf{x} + \mathbf{w} \in \mathbb{C}^{N_R \times 1},
\label{eq:received_signal} \vspace{-0.5ex}
\end{equation}
where $\mathbf{H} \in \mathbb{C}^{N_R \times N_T}$ is {\color{black} the wireless channel matrix} between the receiver and transmitter antennas, and $\mathbf{w} \in \mathbb{C}^{N_R \times 1}$ is the \ac{AWGN} vector with \ac{i.i.d.} elements $w_{n} \sim \mathcal{CN}(0, N_0)$ for $n \in \{1,\cdots\!,N_R\}$ {\color{black} where $N_0$ is the noise power}.

The complex-valued system model in eq. \eqref{eq:received_signal} can be transformed into an equivalent system in the real domain as

\quad\\[-5ex]
\begin{subequations}
\label{eq:IQ_received_signal}
\begin{eqnarray}
&\bm{y} \triangleq 
\begin{bmatrix}
\Re\{\mathbf{y}\} \\
\Im\{\mathbf{y}\} \\
\end{bmatrix}
\!\!\!&=
\overbrace{\begin{bmatrix}
\Re\{\mathbf{H}\} & \!\!\!\!-\Im\{\mathbf{H}\}\\
\Im\{\mathbf{H}\} & \Re\{\mathbf{H}\} \\
\end{bmatrix}}^{\triangleq \;\! \bm{H}} 
\overbrace{
\begin{bmatrix}
\mathbf{x}\R \\
\mathbf{x}\I \\
\end{bmatrix}}^{\triangleq \;\! \bm{x}}
+
\overbrace{
\begin{bmatrix}
\Re\{\mathbf{w}\} \\
\Im\{\mathbf{w}\} \\
\end{bmatrix}}^{\triangleq \;\! \bm{w}}, \label{eq:IQ_received_signal_bivariate} ~~~~~\\
& & = \bm{H}\bm{x} + \bm{w} \in \mathbb{R}^{2N_R \times 1},
\label{eq:IQ_received_signal_univariate}
\end{eqnarray}
where $\bm{y} \in \mathbb{R}^{2N_R\times 1}$, $\bm{H} \in \mathbb{R}^{2N_R \times 2N_T}$, $\bm{x} \in \mathbb{R}^{2N_T \times 1}$, and $\bm{w} \in \mathbb{R}^{2N_R \times 1}$ respectively denote the \ac{IQ}-decoupled counterparts of $\mathbf{y}$, $\mathbf{H}$, $\mathbf{x}$, and $\mathbf{w}$; $\bm{H}\triangleq [\bm{H}\R, \bm{H}\I]$ with $\bm{H}\R \in \mathbb{R}^{2N_R \times N_T}$; and $\bm{H}\I \in \mathbb{R}^{2N_R \times N_T}$ is introduced to denote the effective channel components for $\mathbf{x}\R$ and $\mathbf{x}\I$, respectively.
\end{subequations}

The \ac{IQ}-decoupled system in eq. \eqref{eq:IQ_received_signal} is the basis of most \ac{SotA} \ac{QSM} decoders, which estimates the effective transmit signal $\bm{x}$, in knowledge of $\bm{y}$ and $\bm{H}$.
While the linear recovery problem appears trivial, the challenge lies in the infeasible size of the discrete domain $\mathcal{X}$ of $\bm{x}\in \mathbb{R}^{2N_T \times 1}$, with cardinality $Q \!\triangleq\!\! \big(\big\lfloor\!\binom{N_T}{P}\!\big\rfloor_{2}\big)^{\!2} \!\!\cdot\! M^P\!,$ where $\lfloor \cdot \rfloor_2 \!\triangleq\! 2^{\lfloor \log_2(\cdot) \rfloor}$ is the flooring operation to the nearest power of 2.

As can been seen from the {\color{black}cardinality of $\mathcal{X}$}, the codebook size $Q$ scales at a geometric rate on $P$, and at a squared-factorial rate on $N_T$, such that complexity becomes prohibitive even for moderately large \ac{MIMO} scenarios, which is why simulation results in the literature exist only up to $N_T \leq 10$, $P \leq 3$, even with lowest-complexity methods \cite{Rou_TWC22, Rou_Asilomar22_QSM, An_TVT22}.

\section{Proposed Decoupled Vector GaBP Decoder}

In light of the above, this article provides a solution to the combinatorial spatial domain search challenge, by first considering the {\color{black}\ul{piloted}} \ac{GQSM} scenario, where all symbol component values are known at the receiver\footnote{The natural subsequent extension to the full \ac{GQSM} with unknown transmit symbols will be addressed in an upcoming journal article.}.
The pilot symbols are assumed to be arbitrary and can be utilized for other functionalities such as authentication, radar, channel estimation, etc.
However, even with the known pilot symbols, the challenge remains in estimating the unknown indices of the symbols in the combinatorial space, as seen in eq. \eqref{eq:IQ_GQSM_parts}.

\subsection{Signal Reformulation - Unit Vector Decomposition (\acs{UVD})}

First, noticing that each symbol component $s\R_p$ and $s\I_p$ occupies only a single position in the transmit vector (\textit{i.e.,} only transmitted from a single antenna element), the \ac{GQSM} transmit signal described by eq. \eqref{eq:GQSM_transmit_signal} and eq. \eqref{eq:IQ_GQSM_parts} can be more intuitively rewritten as a superposition of the symbol components multiplied by \textit{activation} vectors, \textit{i.e.,}
\begin{equation}
\mathbf{x} = \underbrace{\mathsmaller\sum_{p=1}^{P} s\R_p \mathbf{e}_{k\R_p}}_{\triangleq \;\mathbf{x}\R}  + j\underbrace{\mathsmaller\sum_{p=1}^{P} s\I_p \mathbf{e}_{k\I_p}}_{\triangleq \;\mathbf{x}\I} \in \mathbb{C}^{N_T \times 1},
\label{eq:reformulated_transmit_signal}
\vspace{-0.5ex}
\end{equation}
where an activation vector $\mathbf{e}_t$ is the $t$-th column of a $N_T \times N_T$ identity matrix, which therefore, is a unit vector.

In light of the reformulation in eq. \eqref{eq:reformulated_transmit_signal}, the \ac{IQ}-decoupled received signal vector in eq. \eqref{eq:IQ_received_signal} becomes
\begin{equation}
\bm{y} = \bm{H}
\begin{bmatrix}
\sum_{p=1}^{P} s\R_p \mathbf{e}_{k\R_p} \\
\!\sum_{p=1}^{P} s\I_p \mathbf{e}_{k\I_p} \\
\end{bmatrix}
+
\bm{w} \in \mathbb{R}^{2N_R \times 1},
\label{eq:reformulated_IQ_received_signal}
\end{equation}
where the linear recovery of $\bm{x}$ has been transformed into the \ul{joint estimation} problem of $2P$ activation (unit) vectors.
\newpage

\begin{figure}[H]
\centering
\begin{subfigure}[b]{\columnwidth}
\centering
\includegraphics[width=0.6\columnwidth]{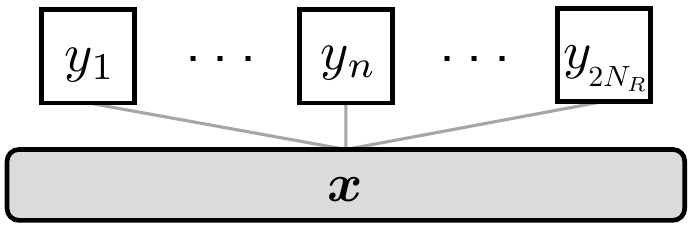}
\vspace{-0.5ex}
\caption{Non-decoupled univariate model in eq. \eqref{eq:IQ_received_signal_univariate} as the \ac{SotA} \cite{Rou_TWC22, An_TVT22, Wang_TVT20, Li_WPC17}.}
\label{fig:factor_graph_single}
\vspace{0.5ex}
\end{subfigure}
\begin{subfigure}[b]{\columnwidth}
\centering
\includegraphics[width=0.6\columnwidth]{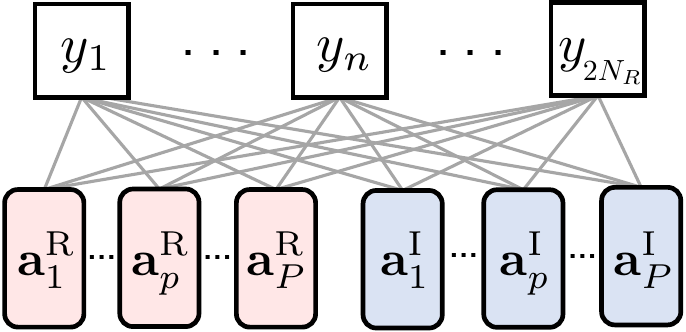}
\vspace{-0.5ex}
\caption{Proposed fully-decoupled $2P$-multivariate \ac{UVD} model in {\color{black}eq. \eqref{eq:RV_received_signal}}.}
\label{fig:factor_graph_proposed}
\end{subfigure}
\vspace{-3.5ex}
\caption{Different \acs{FG} representations of the \ac{GQSM} system.}
\label{fig:factor_graphs}
\vspace{-1.75ex}
\end{figure}

In light of the above, the random vector variable $\mathsf{a}$ is introduced to model the unit vectors, where the discrete uniform prior \ac{PMF} is given by \vspace{-1.5ex}
\begin{equation}
\mathbb{P}_\mathsf{a}(\mathbf{a}) \triangleq \frac{1}{N_T} \cdot \!\sum_{\mathbf{e}_t \in \mathcal{A}} \!\delta(\mathbf{e}_t - \mathbf{a}),
\label{eq:RV_prior_pmf} \vspace{-1.25ex}
\end{equation}
where $\mathbf{a}$ denotes an instance of $\mathsf{a}$, $\mathcal{A} \triangleq \{\mathbf{e}_t\}_{t=1}^{N_T} \in [0,1]^{N_T \times 1}$ is the event set of $\mathsf{a}$, and $\delta(\cdot)$ denotes the unit impulse function where $\delta(\bm{x}) = 1$ if $|\bm{x}|_0 = 0$, and $\delta(\bm{x}) = 0$ otherwise.

Since the $2P$ vector variables are instances of the variable $\mathsf{a}$, the estimation problem is rewritten into the \ac{UVD} form as \vspace{-0.5ex}
\begin{equation}
\bm{y} = \bm{H}
\begin{bmatrix}
\sum_{p=1}^{P} s\R_p \mathbf{a}\R_p \\
\!\sum_{p=1}^{P} s\I_p \mathbf{a}\I_p \\
\end{bmatrix}
+
\bm{w} \in \mathbb{R}^{2N_R \times 1},
\label{eq:RV_received_signal}
\vspace{-0.5ex}
\end{equation}
where the random variables $\mathbf{a}\R_1, \cdots, \mathbf{a}\R_P$ respectively model the unit vectors $\mathbf{e}_{k\R_1}, \cdots, \mathbf{e}_{k\R_P}$ and likewise for $\mathbf{a}\I_1, \cdots, \mathbf{a}\I_P$, which has been illustrated as a \acf{FG} in Fig. \ref{fig:factor_graph_proposed}.


%

\subsection{Vector-valued Gaussian Belief Propagation (\acs{GaBP})}

In light of the above, this section provides the derivation of the purpose-fit vector-valued \ac{MP} rules operating on the factor graph of Fig. \ref{fig:factor_graph_proposed}, based on the \ac{GaBP} framework \cite{Bickson_ARX08,Rou_Asilomar22_QSM}, {\color{black} assuming perfect \ac{CSI} at the receiver.}
This enables the \ul{joint} estimation of the $2P$ activation vector variables (unit vectors) only within their respective signal domains of size $N_T$ each.

First, the soft-replica vectors for the activation vector variables $\mathbf{a}\R_p$ and $\mathbf{a}\I_p$ for $p \in \{1, \cdots, P\}$ are defined as $\hat{\mathbf{a}}\R_{p:n}$ and $\hat{\mathbf{a}}\I_{p:n}$ respectively for the $n$-th factor node with $n \in \{1, \cdots, 2N_R\}$.
The corresponding expected error covariance matrix of the soft-replica $\hat{\mathbf{a}}\R_{p:n}$ is defined as \vspace{-0.75ex}
\begin{equation}
\mathbf{\Gamma}\R_{p:n} \!\triangleq \Exp_{\mathsf{a}}[(\mathbf{a} - \hat{\mathbf{a}}\R_{p:n})(\mathbf{a} - \hat{\mathbf{a}}\R_{p:n})\transs].
\label{eq:error_covariance}
\end{equation}  
\vspace{-3ex}

\noindent \textit{\textbf{Remark:}} \textit{Due to page limitations, the derivations are provided \ul{only for the real components} (i.e., for $\mathbf{a}\R_p$), as the expressions for the respective imaginary components are \ul{identical}, except for the change of superscripts $(\cdot)\R$ to $(\cdot)\I$ and vice-versa.}
\vspace{0.25ex}

In hand of the soft-replicas, the factor nodes perform soft-\ac{IC} on the received signals $y_n$ as \vspace{-2ex}
\begin{equation}
\bar{y}\R_{p:n} = y_n \!-\! {\mathbf{h}_n\R}\transs \! \textstyle\sum_{p' \neq p}^{P} (s_{p'}\R \hat{\mathbf{a}}_{p':n}\R) - {\mathbf{h}_n\I}\transss \! \textstyle\sum_{p' = 1}^{P} (s_{p'}\I \hat{\mathbf{a}}_{p':n}\I)
\vspace{-2.2ex} \label{eq:soft-ic}
\end{equation} \vspace{-1.5ex}
\begin{align}
&\hspace{-0.35ex}=\! \underbrace{{\mathbf{h}_n\R}\transs \!\! s_{p}\R \mathbf{a}_{p}\R}_{\text{true symbol}} \!+ {\mathbf{h}_n\R}\transs\!\! \!\textstyle\sum\limits_{p' \neq p}^{P} \!\!\!s_{p'}\R \!(\;\!\!\hat{\mathbf{a}}_{p'}\R \!\!-\! \hat{\mathbf{a}}_{p'\!:n}\R\!\;\!) \!+\! {\mathbf{h}_n\I}\transss\!\!\! \textstyle\sum\limits_{p' = 1}^{P} \!\!\!s_{p'}\I\! (\;\!\!\hat{\mathbf{a}}_{p'}\I \!\!-\! \hat{\mathbf{a}}_{p'\!:n}\I\!\!\;) \!+\! w_n, \nonumber \vspace{-15ex}
\end{align}

\newpage 

\noindent where ${\mathbf{h}_n\R}\transs\! \!\in\! \mathbb{R}^{1 \times N_T}$ and ${\mathbf{h}_n\I}\transss \!\!\in\! \mathbb{R}^{1 \times N_T}$ respectively denote the $n$-th rows of the channel components $\bm{H}\R \in \mathbb{R}^{2N_T \times N_T}$ and $\bm{H}\I \in \mathbb{R}^{2N_T \times N_T}$ which are defined from $\bm{H} \triangleq [\bm{H}\R, \bm{H}\I]$.

The sum of the latter error terms and \ac{AWGN} term $w_n$, excluding the true symbol part, is approximated as a Gaussian scalar via the \ac{CLT}, which yield the conditional \acp{PDF} of the soft-\ac{IC} symbols with respect to a given activation vector $\mathbf{a}\R_p$ as
\vspace{-1.5ex}
\begin{equation}
\mathbb{P}(\bar{y}_{p:n}\R|\mathbf{a}_{p}\R) \propto \exp\!\bigg( \!-\! \frac{|\bar{y}_{p:n}\R - s_{p}\R {\mathbf{h}_n\R}\transs\! \mathbf{a}_{p}\R|^2}{\nu\R_{p:n}} \bigg),
\label{eq:conditional_pdf}
\vspace{-1ex}
\end{equation}
where the conditional variance $\nu\R_{p:n}$ is obtained by \vspace{-1ex}
\begin{align}
\label{eq:conditional_variances}
\nu\R_{p:n} & = \Exp\!\left[ |\bar{y}_{p:n}\R - s_{p}\R {\mathbf{h}_n\R}\transs\! \mathbf{a}_{p}\R|^2 \right] \nonumber \\[-0.5ex]
& = \nu_{n} - {\mathbf{h}_n\R}\transs\! \big( |s_{p}\R|^2 \cdot \mathbf{\Gamma}_{p:n}\R \big) (\mathbf{h}_n\R)^* + \tfrac{N_0}{2},
\end{align}

\vspace{-1.5ex}
\noindent with $\nu_{n}\!\triangleq\!{\mathbf{h}_n\R}\transs \!\big(\!\sum\limits_{p = 1}^{P}\!\!|s_{p}\R|^2\!  \cdot \!\mathbf{\Gamma}_{p:n}\R \big) {\mathbf{h}_n\R}^{\!*} \!\! +\!
{\mathbf{h}_n\I}\transss\!\big(\!\sum\limits_{p = 1}^{P}\! \!|s_{p}\I|^2 \!\cdot\! \mathbf{\Gamma}_{p:n}\I \big) {\mathbf{h}_n\I\!}^*\!.$

Then, each variable node aggregates the conditional \acp{PDF} from the connected factor nodes to compute the extrinsic belief $b_{p:n}\R$ with self-interference cancellation, following \vspace{-1ex}
\begin{equation}
\mathbb{P}(b_{p:n}\R|\mathbf{a}_{p}\R) = \!\!\prod_{n' \neq n}^{2N_R} \!\!\mathbb{P}(\bar{y}_{p:n'}\R|\mathbf{a}_{p}\R) \label{eq:extrisinc_pdf} \propto e^{({\bm{\eta}\R_{p:n}\!\!\!\!}\trans \mathbf{a}_{p}\R - \tfrac{1}{2}{\mathbf{a}_{p}\R}\trans \mathbf{\Lambda}\R_{p:n} \mathbf{a}_{p}\R)}, \vspace{-1ex}
\end{equation}   
where $\bm{\eta}_{p:n}\R$ and $\mathbf{\Lambda}_{p:n}\R$ are the information vector and the precision matrix of the extrinsic belief $b_{p:n}\R$, given by
\vspace{-1ex}
\begin{equation}
\label{eq:information_vector}
\bm{\eta}_{p:n}\R\! =\! s\R_{p} \!\sum_{n' \neq n}^{2N_R} \!\frac{\bar{y}\R_{p:n'}}{\nu\R_{p:n'}}\mathbf{h}_{n'}\R
\;\text{and}\;
\mathbf{\Lambda}_{p:n}\R\! =\! |s_p\R|^2 \!\! \sum_{n' \neq n}^{2N_R} \!\frac{\mathbf{h}_{n'}\R {\mathbf{h}_{n'}\R}\transss\!}{\nu\R_{p:{n'}}}.
\vspace{-1ex}
\end{equation}

In turn, the posterior Bayes-optimal soft-replicas are computed from the extrinsic beliefs via \vspace{-0.5ex}
\begin{equation}
\hspace{-1ex}\hat{\mathbf{a}}_{p:n}\R  \!=\! \dfrac{\Exp_\mathsf{a}[\mathbf{a} \cdot \mathbb{P}(b_{p:n}\R|\mathbf{a})]}{\Exp_\mathsf{a}[\mathbb{P}(b_{p:n}\R|\mathbf{a})]} \!=\! \dfrac{\sum_{q=1}^{N_T} \mathbf{a}_q \cdot \mathbb{P}(b_{p:n}\R|\mathbf{a}_q) \cdot \mathbb{P}(\mathbf{a}_q)}{\sum_{q'=1}^{N_T} \mathbb{P}(b_{p:n}\R|\mathbf{a}_{q'}) \cdot \mathbb{P}(\mathbf{a}_{q'})},\!\!\!\!
\label{eq:posterior_soft-replica} \vspace{-0.5ex}
\end{equation}

\noindent while the corresponding error covariance matrix is given by \vspace{-0.5ex}
%
\begin{align}
\mathbf{\Gamma}_{p:n}\R &\triangleq \Exp_{{\mathsf{a}}}[(\mathbf{a} - {\hat{\mathbf{a}}}_{p:n}\R)(\mathbf{a}_p - \hat{{\mathbf{a}}}_{p:n}\R)\herm].\label{eq:posterior_error_covariance}
%
%
%
\end{align}
%

%
%
%
%

\vspace{-0.5ex}

Equations \eqref{eq:soft-ic}-\eqref{eq:posterior_error_covariance} describe the steps of one \ac{MP} iteration to estimate the $2P$ activation vectors of the \ac{GQSM} reformulated as eq. \eqref{eq:RV_received_signal}, which yields the refined posterior soft-replica vectors and the corresponding error covariance matrices.
In addition, at the end of such $\tau$-th \ac{MP} iteration, the soft-replica vectors and the error covariance matrices are updated with damping \cite{Som_ITW10} to prevent an early convergence to a local optima \cite{Su_TSP15,Du_MLR17}, following $x^{[\tau + 1]} \!\leftarrow\! \rho x^{[\tau]} \!+\! (1-\rho) x^{[\tau + 1]}$ where $\rho \in [0,1]$ is the damping factor, and $\tau$ is the iteration number.

Next, to obtain the hard-decisions on the $2P$ activation vectors, a the information is aggregated between all $2N_R$ factor nodes to compute the consensus beliefs (\textit{i.e.,} eq. \eqref{eq:extrisinc_pdf} without the self-interference cancellation), which yields the extrinsic consensus \acp{PDF} of the belief $b_{p}\R$.

Finally, the optimal activation vector estimate is selected by evaluating the \acp{PDF} for the $N_T$ valid states of $\mathbf{a} \in \mathcal{A}$ \textit{i.e.,} \vspace{-1.1ex}
\begin{equation}
\tilde{\mathbf{a}}\R_p = \underset{\mathbf{a} \in \mathcal{A}}{\mathrm{argmax}} ~\mathbb{P}(b_{p}\R|\mathbf{a}).
\label{eq:final_decision}
\vspace{-1ex}
\end{equation}

The proposed \ac{UVD}-\ac{GaBP} decoder for pilotted \ac{GQSM}, described by eq. \eqref{eq:error_covariance}-\eqref{eq:final_decision}, is summarized in Algorithm \ref{alg:DUV-GaBP}.

\newpage
\quad
\vspace{-5ex}
\begin{algorithm}[H]
\hrulefill
\begin{algorithmic}[1]
\vspace{-1ex}
\Statex \hspace{-3ex} {\bf{Inputs:}} Received signal $\bm{y}$, 
effective channels $\bm{H}\R$ and $\bm{H}\I$, 
\Statex \hspace{4.65ex} pilot symbols $s\R_p, s\I_p \;\forall p$, and noise variance $N_0$.
\Statex \hspace{-3ex} {\bf{Outputs:}} Estimated activation vectors $\tilde{\mathbf{a}}\R_p$ and $\tilde{\mathbf{a}}\I_p \;\forall p$.
\vspace{-1.5ex}
\Statex \hspace{-4ex}\hrulefill \vspace{-0.5ex}
\Statex \hspace{-3ex} \textbf{Initialization:} $\forall n$ and $\forall p$,
\State Initialize $\hat{\mathbf{a}}\R_{p:n}, \hat{\mathbf{a}}\I_{p:n}$;
\State Compute $\mathbf{\Gamma}\R_{p:n},\mathbf{\Gamma}\I_{p:n}$ via eq. \eqref{eq:error_covariance};
\Statex \hspace{-3ex} \textbf{MP iterations for} $\tau = 1,\cdots,\tau_\mathrm{max}$, $\;\forall n$ and $\forall p$,
\Statex \hspace{-3ex}\textit{For both real and imaginary components:}
\State Perform soft-\ac{IC} via eq. \eqref{eq:soft-ic};
\State Compute $\nu\R_{p:n}, \nu\I_{p:n}$ via eq. \eqref{eq:conditional_variances};
\State Compute $\bm{\eta}_{p:n}\R, \bm{\eta}_{p:n}\I$ and $\mathbf{\Lambda}_{p:n}\R,\mathbf{\Lambda}_{p:n}\I$ via eq. \eqref{eq:information_vector};
\State Compute $\hat{\mathbf{a}}_{p:n}\R, \hat{\mathbf{a}}_{p:n}\I$ via eq. \eqref{eq:posterior_soft-replica};
\State Compute $\mathbf{\Gamma}_{p:n}\R, \mathbf{\Gamma}_{p:n}\I$ via eq. \eqref{eq:posterior_error_covariance};
\State Update $\hat{\mathbf{a}}\R_{p:n}, \hat{\mathbf{a}}\I_{p:n}$ via a damped update \cite{Som_ITW10,Su_TSP15};
\Statex \hspace{-3ex} \textbf{end for}
\State Obtain $\tilde{\mathbf{a}}\R_p$, $\tilde{\mathbf{a}}\I_p$ via eq. \eqref{eq:final_decision};
\caption[]{\!\!: Proposed \acs{UVD}-\ac{GaBP} \ac{GQSM} Decoder}
\label{alg:DUV-GaBP}
\end{algorithmic}
\vspace{-0.5ex}
\end{algorithm}
\setlength{\textfloatsep}{12pt}
\vspace{-3ex}

\section{Performance Evaluation}
\vspace{-0.5ex}
\subsection{Simulation Results}
\vspace{-0.5ex}

\begin{figure}[t]
\vspace{-1ex}
\centering
\includegraphics[width=1\columnwidth]{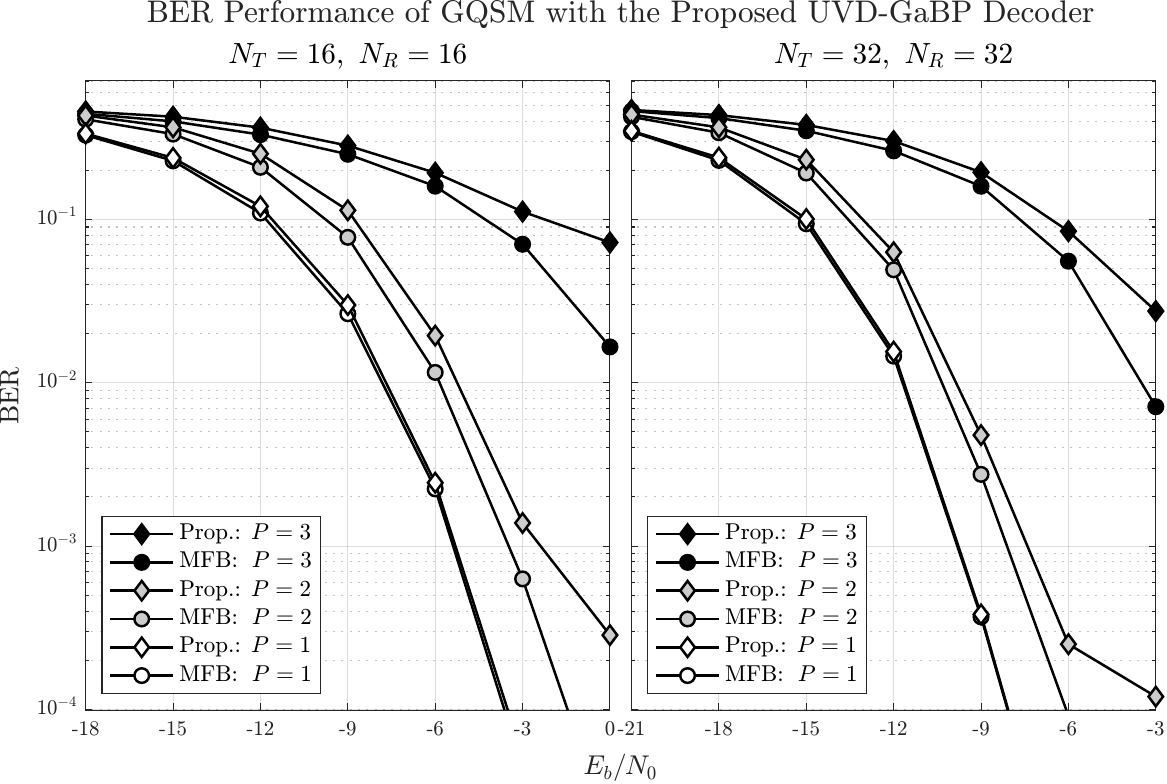}
\vspace{-3.5ex}
\caption{\ac{BER} performance of the proposed (Prop.) \ac{UVD}-\ac{GaBP} decoder and its Genie-aided \ac{MFB}.}
\label{fig:ber_results}
\vspace{-2ex}
\end{figure}

In Fig. \ref{fig:ber_results}, \ac{GQSM} simulation results are provided for \ac{mMIMO} systems with $N_T=16$ and $32$, and varying values of $P$, where the performance is evaluated in terms of the \ac{BER} against the $E_b/N_0$ (\ac{SNR} per bit).

Note the extremely low $E_b/N_0$ ranges of the \ac{GQSM}, which benefits from the fact that most information is encoded without using any transmission power, corroborating the original motivation of enabling energy- and spectral-efficient \ac{mMIMO}.

Since the computational complexity of the brute-force \ac{ML} and \ac{SotA} decoders are prohibitive in the considered system scales, a Genie-aided \acf{MFB} \cite{Takahashi_TC19} is introduced instead as an absolute performance bound, which is obtained by providing the \ac{UVD}-\ac{GaBP} method with perfect prior knowledge of the activation vectors and pilot symbols.

Fig. \ref{fig:ber_results} demonstrates the efficient demodulation capability of the proposed \ac{UVD}-\ac{GaBP} decoder for high-rate \ac{GQSM} signals in $16 \times 16$ and $32 \times 32$ \ac{mMIMO}\footnote{\color{black} Unbalanced \ac{MIMO} scenarios will be investigated in future works.} setups\footnote{Fixed \ac{MP} parametrization have been used for all scenarios, with damping factor $\rho = 0.5$ and number of \ac{MP} iterations $\tau = 100$.}, even with the affordable computational power of an average personal computer.
It can be observed that optimal performance is achieved by the \ac{UVD}-\ac{GaBP} for $P = 1$ in both scenarios, and a slight performance loss of about $1\!\sim\!2 \mathrm{dB}$ and an error-floor is exhibited at high $E_b/N_0$, for increasing $P \geq 1$.
However, notice that the negative effect in both the \ac{BER} performance and the error-floor is reduced in the larger system with $N_T = 32$, which benefits from the increased sparsity in the system and the consequently increased orthogonality in the unit vector random variables.

Further improvements including the elimination of error-floors can be expected if the joint distributions of the activation vectors and their non-uniform prior distributions are introduced in the \ac{MP} design, and a non-piloted variation of the method is also under investigation, which will be addressed in an upcoming journal article. \\[-5ex]

\subsection{Complexity Analysis}

Table \ref{tab:complexity_comparison} compares the decoding complexity of the proposed \ac{UVD}-\ac{GaBP} method against few \ac{SotA} algorithms\footnote{For fairness, the complexity of the symbol-level detection have been disregarded for the \ac{SotA} methods, since a fully pilotted scenario is considered.}, where it can be seen that the \ac{SotA} methods have reduced the squared-combinatorial term appearing in the brute-force \ac{ML} search.

Namely, the \ac{ROMP-OSIC} decoder proposed in \cite{An_TVT22} relaxes the upper index of the binomial coefficient by a factor $N_v$ with $P \!\leq\! N_v \!\leq\! N_T$, whereas the \ac{IQ}-decoupled \ac{GaBP} decoder proposed in \cite{Rou_Asilomar22_QSM} eliminates the quadratic factor on the binomial coefficient, and $\tau_\mathrm{max}$ is the number of \ac{MP} iterations.
However, the \ac{SotA} methods still retain the binomial coefficient which is not scalable to \ac{mMIMO} system of consideration.

On the other hand, the proposed \ac{UVD}-\ac{GaBP} decoder enjoys a significantly reduced complexity\footnote{Since the variables $\mathbf{a}_p$ are unit vectors, the corresponding soft-replicas and covariance matrices reach high sparsity at convergence, such that the practical computational complexity is further reduced with each \ac{MP} iteration.} which is completely independent of the binomial coefficient, enabling the decodability of \ac{GQSM} schemes in significantly larger \ac{mMIMO} systems, as verified in the performance evaluation results.

\begin{table}[H]
\vspace{-1.5ex}
\centering
\normalfont
\caption{Complexity orders of various \ac{GQSM} decoders.}
\vspace{-1ex}
\begin{tabular}{|c||c|}
\hline 
\ac{GQSM} Decoding Algorithm & Decoding Complexity Order \\
\hline\hline
& \\[-2.2ex]
Brute-force \ac{ML} Search \cite{Mesleh_TVT14} & $\mathcal{O}\big[ \binom{N_T}{P}{}^{\!2} P  N_R \big]$ \\ 
& \\[-2.1ex] \hline
& \\[-2.1ex]
\acs{ROMP-OSIC} \cite{An_TVT22} & $\mathcal{O}\big[ \binom{N_v}{P}{}^{\!2} {N_v}^{2}  P N_T N_R \big]$\\
& \\[-2.1ex] \hline
& \\[-2.1ex]
\ac{IQ}-decoupled \ac{GaBP} \cite{Rou_Asilomar22_QSM} & $ \mathcal{O}\big[ \;\!\tau_\mathrm{max} \!\cdot\! \binom{N_T}{P} {N_T}^{2} N_R \big]$ \\  
& \\[-2.1ex] \hline 
& \\[-2ex]
Proposed \ac{UVD}-\ac{GaBP} & $\mathcal{O}\big[ \;\!\tau_\mathrm{max} \!\cdot\! P {N_T}^{2} N_R \big]$ \\[-2.2ex] 
& \\ \hline \hline
\end{tabular}
\label{tab:complexity_comparison}
\end{table}

\vspace{-2ex}
\section{Conclusion}
\vspace{-0.5ex}

We paved the way towards feasible energy- and spectral-efficient \ac{mMIMO} systems with high-performance \ac{GQSM}, by proposing a novel \ac{GaBP}-based decoder exploiting a \ac{UVD} of the \ac{GQSM} signal model and pilots, which is shown to achieve a complexity order that is independent of combinatorial factors.
Simulation results verify the effectiveness of the method.

{\color{black} In addition, the multi-user scenario should also be considered to support the \ac{B5G} \ac{mMIMO} access expectations.}
\vspace{-1ex}


\newpage
\bibliographystyle{IEEEtran}
\bibliography{listrefs_HSRou}

\end{document}